\documentclass[a4paper]{jpconf}
\usepackage{graphicx}
\begin{document}
\title{Axions and the white dwarf luminosity function}

\author{J Isern$^{1,2}$, 
        S Catal\'an$^{1,2}$, 
        E Garc\'{\i}a--Berro$^{2,3}$ and 
        S Torres$^{2,3}$ }

\address{$^1$Institut de Ci\`encies de l'Espai (CSIC), 
             Facultat de Ci\`encies, 
             Campus UAB, 
             08193 Bellaterra, 
             Spain}
\address{$^3$Institut d'Estudis Espacials de Catalunya (IEEC), 
             c/ Gran Capit\`{a} 2--4, 
             08034 Barcelona, 
             Spain} 
\address{$^3$Departament de F\'isica Aplicada, 
             Escola Polit\`ecnica Superior de Castelldefels,
             Universitat Polit\`ecnica de Catalunya, 
             Avda. del Canal Ol\'\i mpic s/n, 
             08860 Castelldefels, 
             Spain}

\ead{isern@ieec.cat, catalan@ieec.cat, garcia@fa.upc.edu, santi@fa.upc.edu}

\begin{abstract}
The evolution  of white  dwarfs can be  described as a  simple cooling
process.   Recently,  it  has  been  possible  to  determine  with  an
unprecedented precision their luminosity function, that is, the number
of stars per unit volume  and luminosity interval.  Since the shape of
the bright  branch of this function  is only sensitive  to the average
cooling rate, we use this  property to check the possible existence of
axions, a  proposed but not yet detected  weakly interacting particle.
We  show  here that  the  inclusion of  the  axion  emissivity in  the
evolutionary models of white  dwarfs noticeably improves the agreement
between the theoretical calculations and the observational white dwarf
luminosity function, thus providing the first positive indication that
axions  could  exist.  Our  results  indicate  that  the best  fit  is
obtained for $m_{\rm  a} cos^2 \beta \simeq 2 -  6$ meV, where $m_{\rm
a}$ is the mass of the axion and $cos^2\beta$ is a free parameter, and
that values larger than 10 meV are clearly excluded.
\end{abstract}

\section{Introduction}

White   dwarfs  are  the   final  evolutionary   stage  of   low-  and
intermediate-mass  stars ($M\leq  10\pm 2  M_\odot$).  Since  they are
degenerate  objects,  they  cannot  obtain energy  from  thermonuclear
reactions  and  their evolution  is  just  a  gravothermal process  of
cooling.   They  have a  relatively  simple  structure  composed by  a
degenerate core that contains the bulk of mass and acts as a reservoir
of  energy, and  a  partially degenerate  envelope  that controls  the
energy outflow.   White dwarfs  masses are in  the range of  $0.3 \leq
M/M_\odot \leq 1.4$ (the  Chandrasekhar's mass). Those with $M\leq 0.4
M_\odot$  have a core  made of  He, while  those with  a $M  \geq 1.05
M_\odot$ have  a core made of O  and Ne. The remaining  ones, the vast
majority, have a  core made of a mixture  of C and O. All  of them are
surrounded by a  thin helium layer with a  mass ranging from $10^{-2}$
to  $10^{-4}\, M_\odot$,  which, in  turn,  is surrounded  by an  even
thinner layer of hydrogen with a mass between $10^{-4}$ and $10^{-15}$
$M_\odot$,  although about  25\%  of  white dwarfs  do  not have  such
hydrogen envelopes.  White dwarfs displaying hydrogen in their spectra
are known  as DA  and the  remainning ones as  non-DA. Because  of the
different  opacities involved,  DA white  dwarfs cool  down  much more
slowly than  non-DA white dwarfs.

Because of the simplicity of their structures, it has been proposed to
use white dwarfs as laboratories  to test new physics. The reasons for
such  a proposal  are the  following ones.  Firstly, the  evolution of
white dwarfs is just a  simple process of cooling. Secondly, the basic
physical  ingredients necessary  to predict  their evolution  are well
identified,  although not necessarily  well understood,  and, finally,
there  is a  solid and  continuously growing  observational background
that  allows  to  test  the  different  theories.   Therefore,  if  an
additional  source or  sink  of energy  is  added, the  characteristic
cooling time is modified and these changes can be detected through the
anomalies  introduced in  the luminosity  function or  in  the secular
drift of the pulsation period of degenerate variables --- see Isern \&
Garc\'{\i}a--Berro  (2008) for details.  In this  paper we  will limit
ourselves  to study  the  changes introduced  by  the introduction  of
axions in the white dwarf luminosity function.

\section{The white dwarf luminosity function}

\begin{figure}[t]
\vskip 7.3cm
\includegraphics{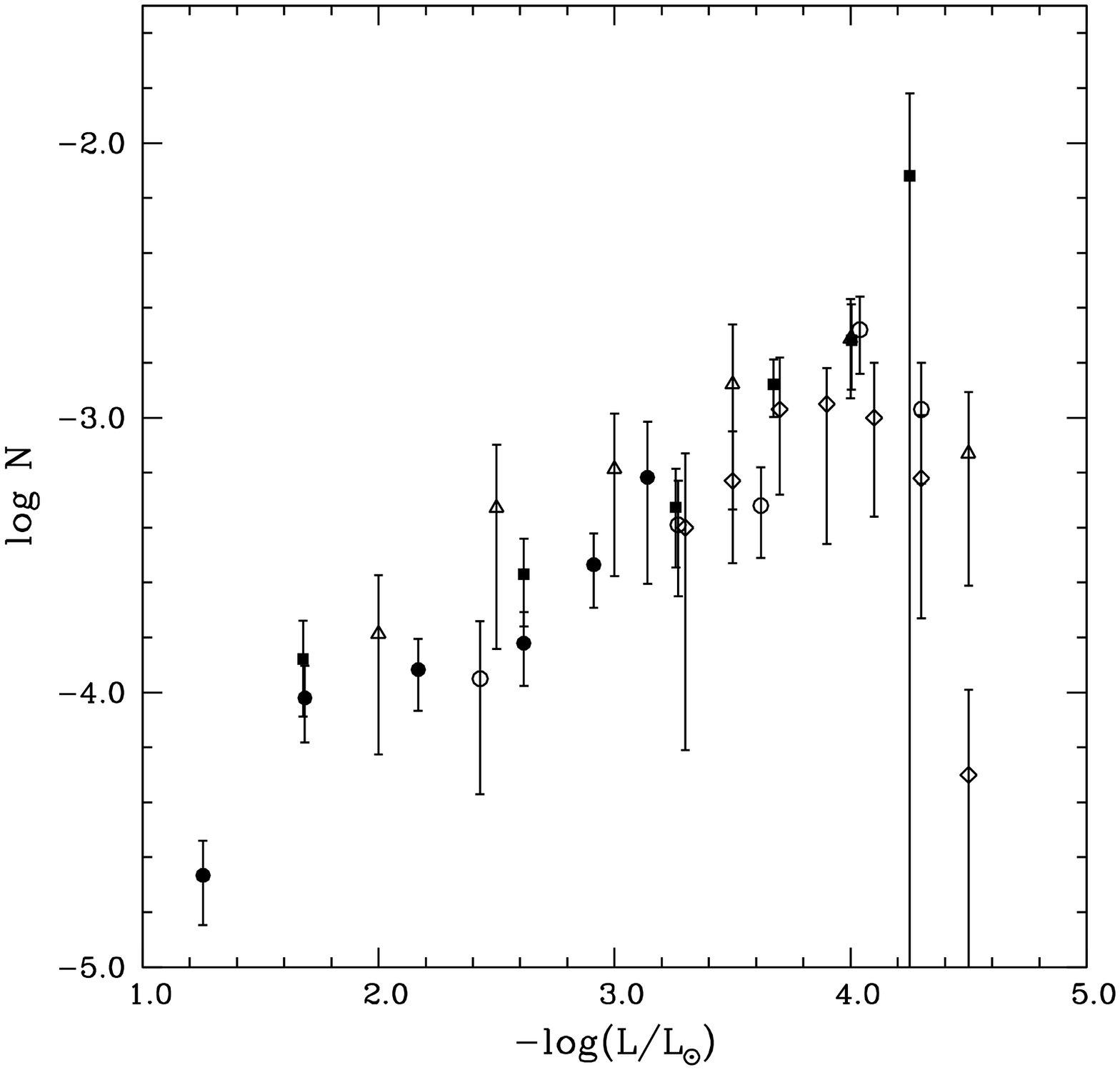}
\includegraphics{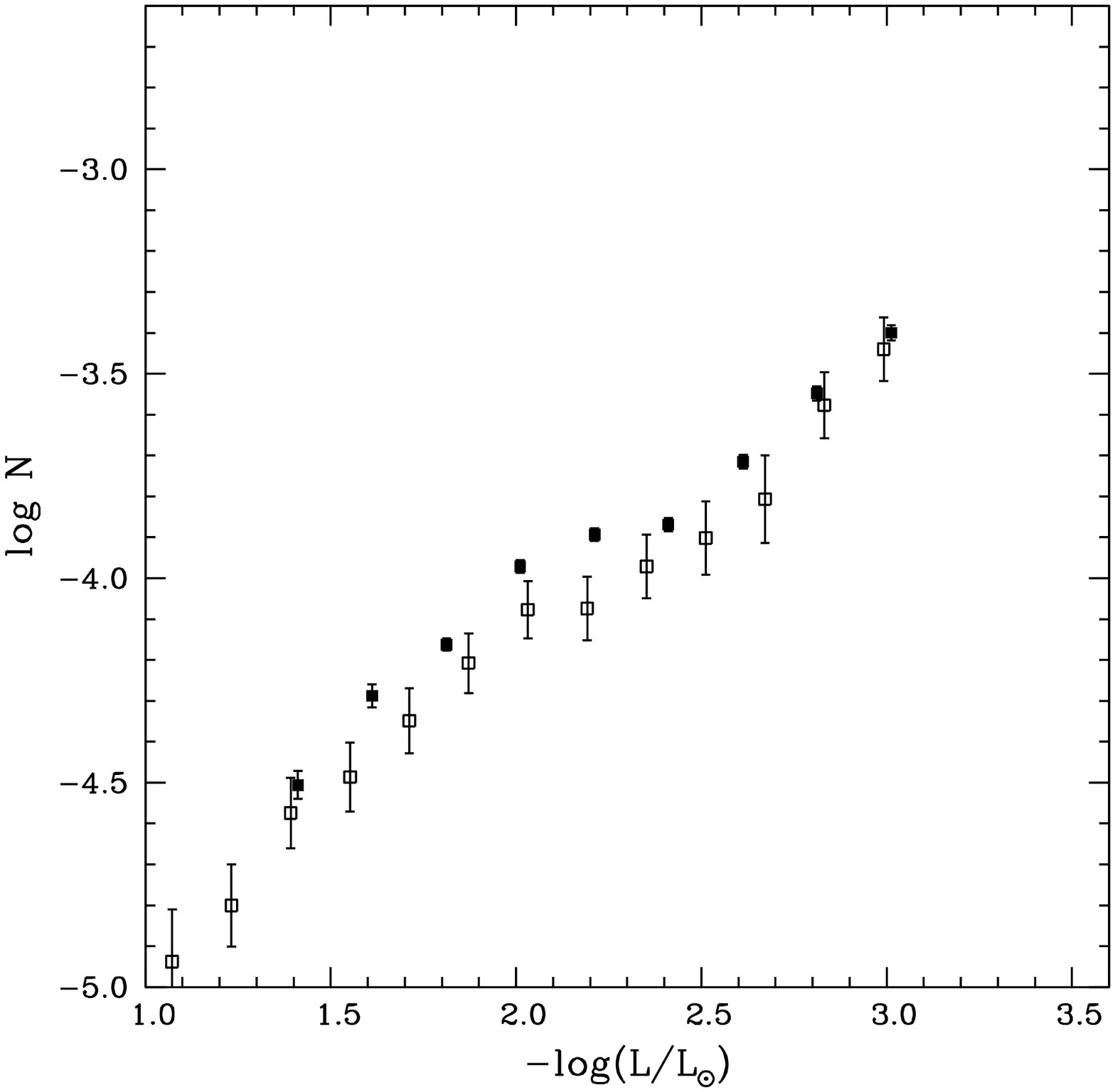}
\caption{Left panel:  luminosity functions obtained before  the era of
         large  surveys.  The  different  symbols represent  different
         determinations: Liebert, Dahn \& Monet (1985), $\fullcircle$;
         Evans (1992),  $\fullsquare$; Oswalt, Smith,  Wood \& Hintzen
         (1996),  $\opentriangle$; Leggett,  Ruiz \&  Bergeron (1998),
         $\opendiamond$;    Knox,    Hawkins    \&   Hambly    (1999),
         $\opencircle$. Right panel: luminosity functions derived from
         the SDSS.  One, $\fullsquare$, is  composed by DA  and non-DA
         white  dwarfs  identified from  their  photometry and  proper
         motion  (Harris et  al. 2008). The  other,  $\opensquare$, is
         only composed by spectroscopically identified DA white dwarfs
         (DeGennaro et al. 2008)}
\label{fig1} 
\end{figure}

The white dwarf  luminosity function is defined as  the number density
of white dwarfs of a given luminosity per unit magnitude interval:

\begin{equation}
n(l) = \int^{M_{\rm s}}_{M_{\rm i}}\,\Phi(M)\,\Psi(\tau)
\tau_{\rm cool}(l,M) \;dM
\label{lf}
\end{equation}
\noindent

\noindent where

\begin{equation}
\tau = T-t_{\rm cool}(l,M)-t_{\rm PS}(M)
\label{bc}
\end{equation}

\noindent and  $l = \log (L/L_\odot)$,  $M$ is the mass  of the parent
star (for  convenience all white dwarfs  are labeled with  the mass of
the main sequence progenitor), $t_{\rm cool}$ is the cooling time down
to   luminosity  $l$,   $\tau_{\rm  cool}=dt/dM_{\rm   bol}$   is  the
characteristic  cooling time,  $M_{\rm  s}$ and  $M_{\rm  i}$ are  the
maximum  and the minimum  masses of  the main  sequence stars  able to
produce a white dwarf of  luminosity $l$, $t_{\rm PS}$ is the lifetime
of  the progenitor  of the  white dwarf,  and $T$  is the  age  of the
population under  study.  The  remaining quantities, the  initial mass
function, $\Phi(M)$,  and the star formation rate,  $\Psi(t)$, are not
known a priori and depend  on the properties of the stellar population
under study. The computed luminosity function is usually normalized to
the bin with the smallest error bar, traditionally the one with $l=3$,
in order to compare theory with observations.  Equation \ref{lf} shows
that in order to use  the luminosity function as a physical laboratory
it is necessary to have good enough observational data and to know the
galactic properties that are used in this equation (the star formation
rate, the initial mass function and the age of the Galaxy).

The first luminosity function was derived four decades ago (Weidemann
1968) and since  then it  has been noticeably  improved, see  the left
panel of  Fig.  \ref{fig1}.  The  monotonic behavior of  this function
clearly proves  that the evolution of  white dwarfs is  just a cooling
process.  The  sharp cut-off at  low luminosities is a  consequence of
the finite age of the Galaxy.  The availability of data from the Sloan
Digital Sky Survey (SDSS) is  noticeably improving the accuracy of the
new luminosity  functions. The one by  Harris (2006) was  built from a
sample of 6000 DA and non-DA white dwarfs with accurate photometry and
proper  motions culled from  the SDSS  Data Release  3 and  the USNO-B
catalogue, whereas the one obtained by DeGennaro et al. (2008) was constructed
from  a sample of  3528 spectroscopically  identified DA  white dwarfs
from  the  SDSS  Data  Release  4  (see  the  right  panel  of  Figure
\ref{fig1}). The quality of the data, specially in the central part of
the bright  branch, is so  good that they  allow to start  testing the
inclusion of new physics in the cooling process of white dwarfs.

\begin{figure}[t]
\vskip 7.3cm
\includegraphics{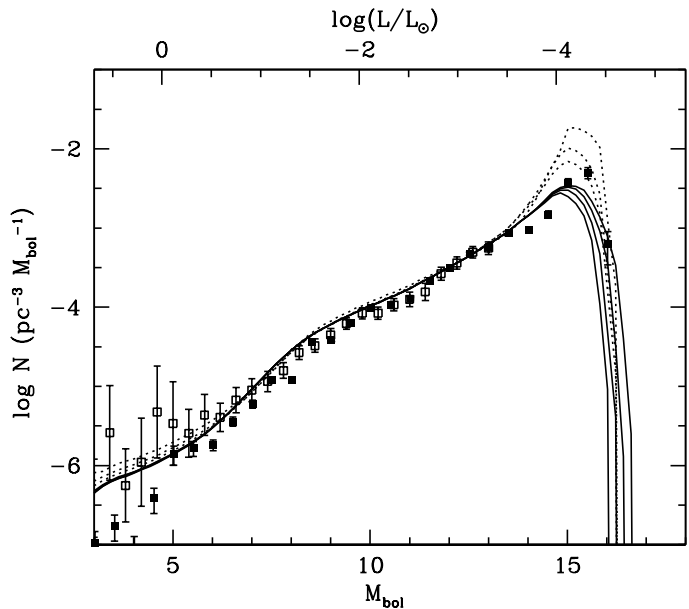}
\caption{Theoretical  luminosity   functions  of  white   dwarfs.  The
         observational  data are  the  same represented  in the  right
         panel  of  figure \ref{fig1}.  Solid  lines  were obtained  for
         different ages  of the  Galaxy --- from  left to
         right:  10,  11,  12 and  13  Gyr  ---  and a  constant  star
         formation rate. Dotted lines were  obtained using an age of 11
         Gyr but  exponentially decreasing star  formation rates, with
         $\tau=0.5$, 3  and 5 Gyr.  All the  luminosity functions have
         been normalized to the same observational data point.}
\label{fig2} 
\end{figure}

It is evident that before  introducing new ingredients it is necessary
to have  a good  model of  cooling able to  reproduce as  accurately as
possible the observations. It is worthwhile here to remember that when
the luminosity  is large, $M_{\rm bol}<8$, the  evolution is dominated
by neutrino emission.  In this  phase the main uncertainties come from
our poor knowledge of the initial conditions. Fortunately, it has been
shown  that all  the  initial thermal  structures  converge towards  a
unique one.  For smaller  luminosities, $8\leq M_{\rm bol}\leq12$, the
main source  of energy is of  gravothermal origin. In  this phase, the
Coulomb plasma coupling parameter is  not large and the cooling can be
accurately  described.   Furthermore,  the  energy  flux  through  the
envelope  is  controlled  by   a  thick  non-degenerate  or  partially
degenerate layer with an  opacity dominated by hydrogen, when present,
and  helium, and it  is weakly  dependent on  the metal  content since
metals    sink    towards    the    base   of    the    envelope    by
gravitationally-induced  diffusion.  Below  these  luminosities, white
dwarfs evolve  into a region  of densities and temperatures  where the
plasma  crystallizes. When  this  happens, two  additional sources  of
energy appear:  the release of latent heat  during crystallization and
the release of gravitational energy induced by phase separation of the
different chemical species (Garc\'\i  a--Berro et al.  1988). When the
bulk  of the  star is  solid, the  white dwarf  enters into  the Debye
cooling phase and  the only important source of  energy comes from the
compression  of the  outer layers  --- see  Isern et  al (1998)  for a
detailed  discussion. These  late phases  of cooling  are not  yet well
understood  (Isern et  al 2000).  Figure \ref{fig2}  shows that  it is
possible to construct a good  standard model with the cooling sequences
of  Salaris et  al.  (2000), the  initial-final  mass relationship  of
Catalan et al. (2008), a  constant star formation rate (solid line) and
an age  of the  disk of 10.5  Gyr.  The  cooling models also  assume a
non-homogeneous distribution of carbon and oxygen in the core (Salaris et al.
1997), a pure helium layer of $10^{-2} \, M_*$ and on top of it a pure
hydrogen  layer of $10^{-4}\,  M_*$, where  $M_*$ is  the mass  of the
white dwarf.

\begin{figure}[t]
\vskip 8cm
\includegraphics{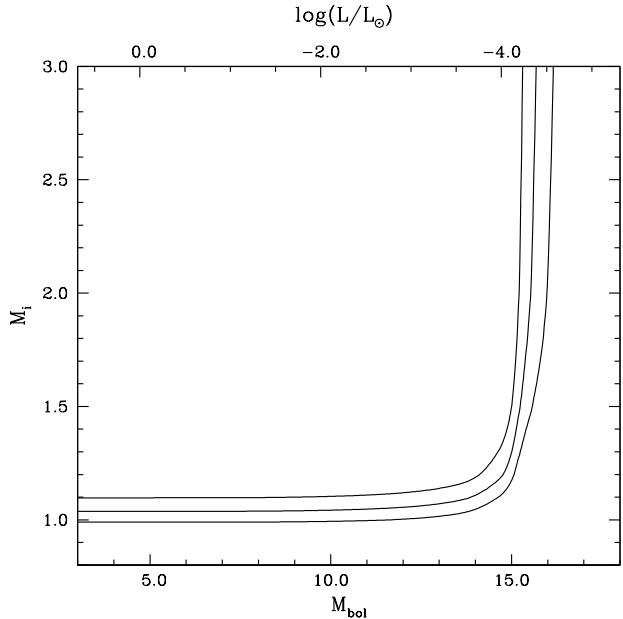}
\caption{Minimum mass of the parent star able to produce a white dwarf
         of luminosity $L$ for different ages of the Galaxy (9, 11 and
         13 Gyr from top to bottom, respectively).}
\label{fig3} 
\end{figure}

The third condition to be fulfilled is the reliability of the Galactic
data: the  age of the Galaxy,  the initial mass function  and the star
formation rate.  An interesting feature  of figure \ref{fig2}  is that
the bright part  of the white dwarf luminosity  function --- that with
bolometric magnitude $M_{\rm bol}<13$ --- is almost independent of the
assumed  star  formation rate.   This  can  be  explained with  simple
arguments.   Since the  characteristic  cooling time  is not  strongly
dependent  on the  mass  of  the white  dwarf,  Eq.~(\ref{lf}) can  be
written as:

\begin{equation}
n = \left\langle\tau_{\rm cool}\right\rangle 
\int_{M_{\rm l}}^{M_{\rm u}}\phi(M)
\psi(T-t_{\rm cool}-t_{\rm ps})\;dM.
\label{lf1} 
\end{equation}

Restricting  ourselves to bright  white dwarfs  --- namely,  those for
which $t_{\rm cool}$  is small --- the lower limit  of the integral is
satisfied  by low-mass  stars  and,  as a  consequence  of the  strong
dependence of the main sequence  lifetimes with mass, it takes a value
that is almost independent  of the luminosity under consideration (see
figure \ref{fig3}). Therefore, if $\psi$  is a well behaved function and
$T_{\rm  G}$ is large  enough, the  integral is  not sensitive  to the
luminosity, its  value is absorbed by the  normalization procedure and
the  shape of  the luminosity  function only  depends on  the averaged
physical properties of  white dwarfs. It is important  to mention here
that the initial-final  mass relationship enters as a  weight into the
calculation of  this average. Neverheless, since  only those functions
able to  provide a good fit  to the mass distribution  of white dwarfs
are acceptable, its influence on the shape of the bright branch of the
luminosity function is minor (Isern et al 2008).

\section{Axions}

One solution to the strong CP problem of quantum chromodynamics is the
Peccei-Quinn symmetry  (Peccei \& Quinn  1977a, b). This  symmetry is
spontaneously  broken  at an  energy  scale  that  gives rise  to  the
formation  of a  light pseudo-scalar  particle named  axion (Weinberg,
1978; Wilczek,  1978). This  scale of energies  is not defined  by the
theory but  it has to  be well above  the electroweak scale  to ensure
that the coupling between axions  and matter is weak enough to account
for the lack  of positive detection up to now. The  mass of axions and
the  energy  scale  are  related  by: $m_{\rm  a}  =  0.6(10^7\,  {\rm
GeV}/f_{\rm  a})$ eV.  Astrophysical and  cosmological  arguments have
been used to  constrain this mass to the range  $10^{-4} {\rm eV} \leq
m_{\rm  a} \leq  10^{-4}{\rm eV}$.   For this  mass range,  axions can
escape from  stars and  act as  a sink of  energy. Therefore,  if they
exist, they can noticeably modify the cooling of white dwarf stars.

\begin{figure}[t]
\vskip 7.3cm
\includegraphics{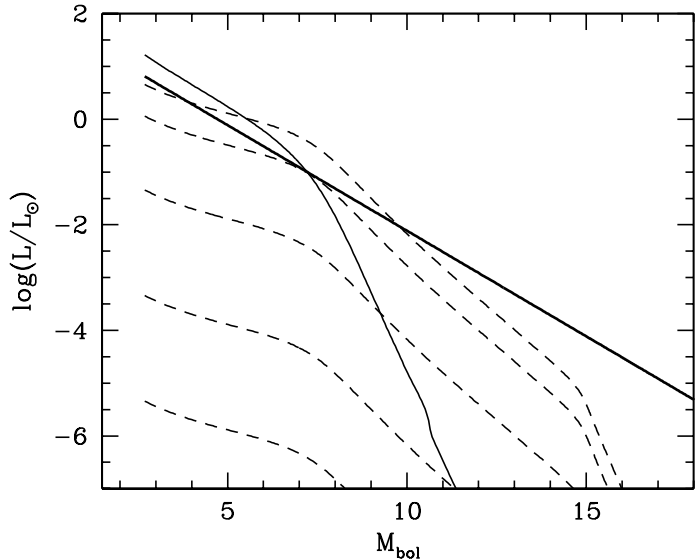}
\caption{Energy losses for a typical $0.61\, M_\odot$ white dwarf as a
         function  of  the  bolometric  magnitude.  The  dashed  lines
         represent  the  axion  luminosity  for  different  values  of
         $m_{\rm a} \cos^2\beta$, where $m_{\rm a}$ is the mass of the
         axion,  and $\cos^2\beta$ is  a free  parameter (from  top to
         bottom: $m_{\rm  a} \cos^2\beta =10,  5, 1, 0.1,  0.01\, {\rm
         meV}$).   The  thick   solid  line   represents   the  photon
         luminosity, while the thin solid line represents the neutrino
         luminosity.}
\label{fig4}
\end{figure}

Axions can couple  to photons, electrons and nucleons  with a strength
that  depends  on  the  specific implementation  of  the  Peccei-Quinn
mechanism.   The two most  common implementations  are the  KSVZ (Kim,
1979; Shifman  et al., 1979) and  the DFSZ models (Dine  et al., 1981;
Zhitniskii,  1980). In  the first  case, axions  couple to  hadrons and
photons, while in the second  they also couple to charged leptons. For
the   temperatures   and  densities   of   the   white  dwarfs   under
consideration, only DFSZ axions are relevant and in this case they can
be emitted by Compton, pair annihilation and bremsstrahlung processes,
but  only  the  last  mechanism  turns out  to  be  important.  Figure
\ref{fig4}  shows the energy  losses for  a typical  white dwarf  as a
function  of the bolometric  magnitude.  The  axion emission  rate (in
erg/g/s)  has   been  computed   (Nakagawa  1987,  1988)   as  $e_{\rm
a}=1.08\times 10^{23}\alpha  Z^2/AT_7^4F$, where $F$ is  a function of
the  temperature  and  the   density  which  takes  into  account  the
properties of the  plasma, $\alpha = g_{\rm ae}^2/4\pi$  is related to
the axion-electron coupling constant  $g_{\rm ae} = 2.8\times 10^{-11}
m_{\rm  a}  \cos^2\beta  /1\,{\rm  eV}$.  Since  the  core  is  almost
isothermal, $L_{\rm a}\propto  T^4$ in the region in  which axions are
the  dominant sink  of energy.  The  thick solid  line represents  the
photon luminosity  (Salaris et  al 2000). For  the region  of interest
$L_\gamma \propto T^\alpha$, with  $\alpha \propto 2.6$, although this
value  changes as  the white  dwarf cools  down. The  thin  solid line
represents  the neutrino  luminosity (Salaris  et al  2000)  --- which
scales  as $L_\nu  \propto T^8$  --- and  is also  dominated  by the
plasma and bremsstrahlung  processes. Therefore, since the temperature
dependence of  the different energy-loss  mechanisms is not  the same,
the   luminosity  function   allows  to   disentangle   the  different
contributions.

\begin{figure}[t]
\vskip 7.3cm
\includegraphics{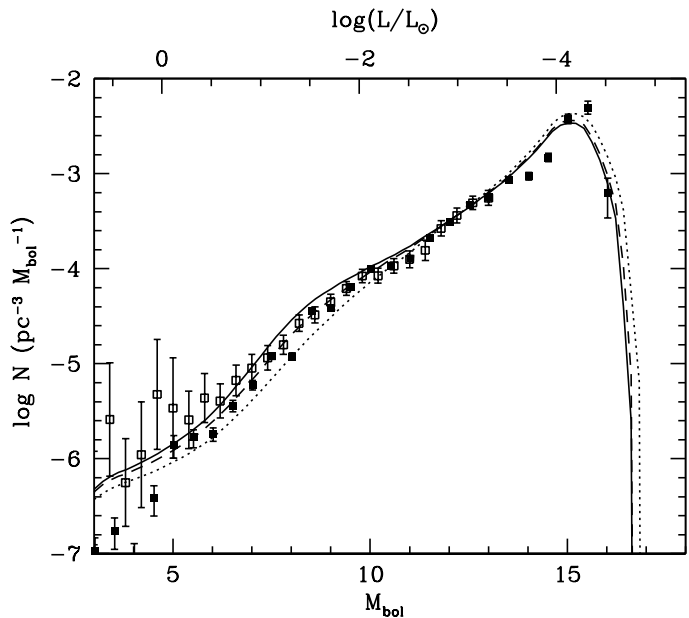}
\caption{White dwarf luminosity  functions for different axion masses:
         $m_{\rm a}\cos^2\beta$ = 0  (solid line), 5 (dashed line) and
         10 (dotted line) meV.}
\label{fig5}
\end{figure}

Figure \ref{fig4} shows  that in the region $M_{\rm  bol} \sim 10$ the
axion luminosity is  not negligible when compared with  the photon and
neutrino ones. It also shows  that the region around $M_{\rm bol} \sim
12$ provides a solid anchor point to normalize the luminosity function
because there the observational data have reasonably small error bars,
models are  reliable, neutrinos are  not relevant and axions,  if they
exist, are not dominant.

\section{Results and conclusions}

\begin{figure}[t]
\vskip 9cm
\includegraphics{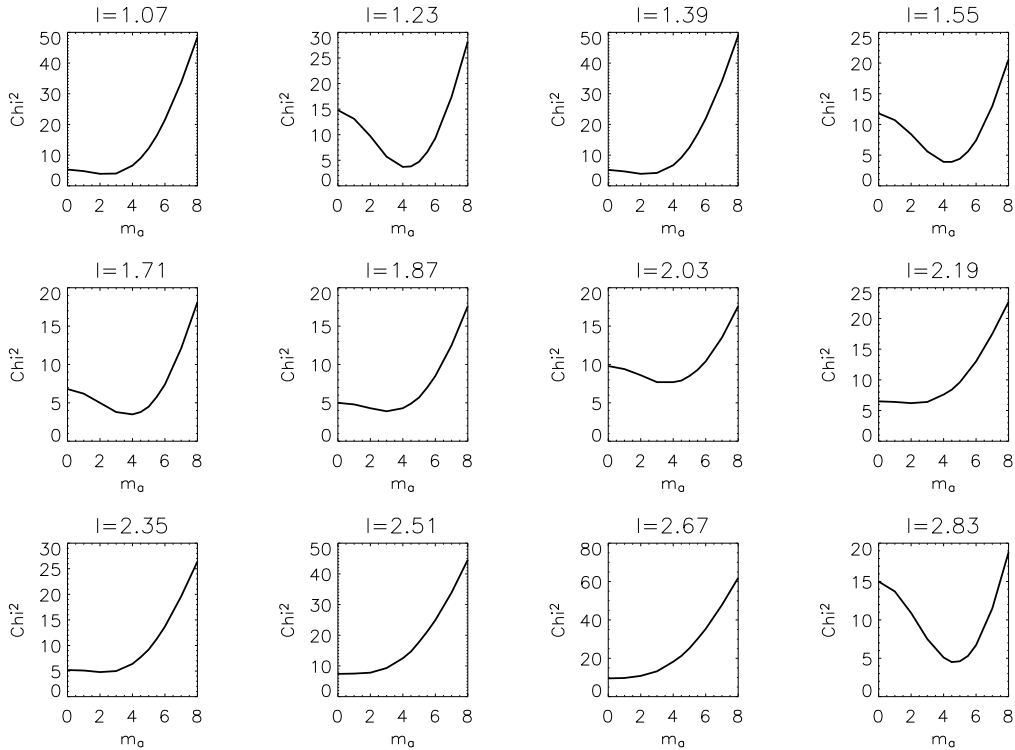}
\caption{Value  of $\chi^2$ as  a function  of the  mass of  the axion
         adopting different normalization values.}
\label{fig6}
\end{figure}

Figure \ref{fig5} displays the luminosity function for
different axion masses, a constant star formation rate and an
age  of the  Galactic disk  of 11  Gyr.  As  already mentioned,  it is
important  to  remember  that  the  bright branch  of  the  luminosity
function  is  not  sensitive  to  these  last  assumptions.   All  the
luminosity  functions have been  normalized to  the luminosity  bin at
$\log  (L/L_{\odot})\simeq-3$ or,  equivalently,  $M_{\rm bol}  \simeq
12.2$.   The  best fit  model  ---  namely  that which  minimizes  the
$\chi^2$ test in the region $-1>\log (L/L_{\odot})>-3$ (that is, $ 7.2
<  M_{\rm  bol}  <  12.2$),   which  is  the  region  where  both  the
observational  data and  the theoretical  models are  reliable  --- is
obtained  for $m_{\rm a}  \cos^2 \beta\approx  5.5$ meV  and solutions
with $m_{\rm a} \cos^2 \beta >  10$ meV are clearly excluded (Isern et
al 2008).   Figure \ref{fig6} displays  the behavior of $\chi^2$  as a
function of the mass of  the axion for different normalization points.
In all cases, $\chi^2$ displays a pronounced minimum for masses in the
range of  roughly 2 to 6 meV  except for $l=2.51$ and  $2.67$ that are
compatible  with $m_{\rm  a} =  0$. However,  as it  can be  seen from
figure  \ref{fig6},  both points  deviate  from  the  behavior of  the
neighbouring values.

Taken  at  face  value,  these  results  not  only  provide  a  strong
constraint to the allowed mass of axions, but also a first evidence of
their  existence and a  rough estimation  of their  mass.  This  is of
course  a strong  statement and  in  order to  be accepted  it has  to
fulfill   the   following  conditions:   it   has   to  be   confirmed
independently, it  has to resist the introduction  of any conventional
effect not previously included, it has to provide testable predictions
and it can not enter  in contradiction with well established facts.  A
detailed discussion of the existing  uncertainties is out of the scope
of the present  paper but it is worthwhile to  enumerate some of them:
influence of  the metallicity in the  age and core  composition of the
progenitor,  transparency of the  envelope and  conversion of  DA into
non-DA white dwarfs, IMF, intial-final mass relationship, pathological
SFRs  or  uncertainties  in  the  observational  luminosity  function,
especially on  the role of He  white dwarfs and the  detailed shape of
the brightest part of this function and others.

The results  found here are  completely compatible with  the presently
known constraints (Raffelt 2007), but since the predicted mass is near
the allowed upper  bound, it is natural to  expect they will introduce
some subtle changes  in the late stages of stellar  evolution as it is
indeed the case  of white dwarfs. A detailed  discussion of this point
is also out of  the scope of this paper and we  will just mention that
these values  are compatible with the  bounds imposed by  the drift of
the  pulsational period  of the  ZZ  Ceti star  G117$-$B15A (Isern  et
al. 1992;  C\'orsico et al. 2001;  Bischoff-Kim et al.   2008).  It is
also worthwhile  to mention  here that axions  with $m_{\rm  a} \cos^2
\beta  \approx  5$ meV  would  change  the  expected period  drift  of
variable DB  white dwarfs --- which  have values between  $\dot P \sim
10^{-13}$ and $10^{-14}$ s~s$^{-1}$ (C\'orsico \& Althaus 2004) --- by
a factor of 2, the exact value depending on the adopted temperature of
the stellar core.   The structure of the Sun would  not be modified by
the  emission of  axions of  this  masses since  they would  represent
$L_{\rm  a} \sim  10^{-6}L_\odot$. However,  an instrument  like CAST,
able to operate in the region  of masses $m_{\rm a}\sim {\rm meV}$ and
coupling constant  $g_{{\rm a}\gamma} \sim  10^{-12}\, {\rm GeV}^{-1}$
could  be able  to detect  them  and confirm  or rule  out the  masses
predicted  here.  Cosmology could  also provide  some insight  to this
problem. Axions have been proposed as dark matter candidates but if we
assume  $\cos^2\beta=1$,  their contribution  would be of  the order  of
$\Omega_{\rm a}  h^2 \sim 10^{-4}$,  where $h$ is the  Hubble constant
normalized to 100 km/s/Mpc (Raffelt  2007), which rules out them as the
main  dark matter  candidate.   On  the contrary,  if  dark energy  is
interpreted  as the  energy density  of vacuum,  axions could  have an
important role. The vacuum energy is obtained from the sum of the zero
point  energy  of all  the  quantum  fields  (where bosons  contribute
positively and  fermions negatively)  and if we  try to  reproduce the
observed value, that is positive,  with just one quantum field we need
a boson  with a mass  of the order  or smaller than 10  meV (Friemann,
Turner \& Huterer,  2008), that is just what we  have found here. This
is indeed a remarkable coincidence.

To  conclude, it  is evident  that our  claim about  the  existence of
axions has  to be taken  with caution, but  it clearly shows  that the
white dwarf luminosity function  can provide strong arguments to solve
the long-lasting problem  of the CP violation.  Future  work should be
devoted  to  improving  the  observational data,  especially  for  the
brightest  part   of  the   white  dwarf  luminosity   function.   The
theoretical models could also  be improved and also additional efforts
should  be   devoted  to  find  new  independent   methods  to  detect
axions. Finally,  it is worthwhile to say  that even in the  case of a
negative result,  an accurate and precise luminosity  function could be
used to provide  insight to many other problems  like the hypothetical
drift  of  the  gravitation  constant  or  the  magnetic  momentum  of
neutrinos just to cite two cases.

\ack 
Part of this  work was supported by the  MEC grants AYA05-08013-C03-01
and 02,  and ESP2007-61593, by the  European Union FEDER  funds and by
the AGAUR.

\section*{References}
\begin{thereferences}
\item Bischoff-Kim A Montgomery M H \& Winget D E 2008 {\sl ApJ} {\bf 675} 1512
\item Catal\'an S Isern J Garc\'{\i}a--Berro E \& Ribas I 2008 {\sl MNRAS} 
      {\bf 387} 1693
\item C\'orsico A H \& Althaus L G 2004 {\sl A\&A} {\bf 428} 159
\item C\'orsico A H Benvenuto O G Althaus L G Isern J \& Garc\'{\i}a--Berro E 
      2001 {\sl New Astron.} {\bf 6} 197 
\item DeGennaro S von Hippel T Winget D E Kepler S O Nitta A Koester D \& 
      Althaus L 2008 {\sl ApJ} {\bf 135} 1
\item Dine M Fishler W \& Srednicki M 1981 {\sl Phys. Lett. B} {\bf 104} 199 
\item Evans D W 1992 {\sl MNRAS} {\bf 255} 521
\item Friemann J A Turner M S \& Huterer D 2008 {\sl ARAA} {\bf 46} 385
\item Garc\'\i a--Berro E Hernanz M Isern J \& Mochkovitch R 1988 {\sl Nature}
      {\bf 333} 642
\item Harris H C, et al. 2006 {\sl AJ} {\bf 131} 571
\item Isern J Mochkovitch R Garc\'{\i}a--Berro E \& Hernanz M 1998 
      {\sl J. Phys.: Cond. Matt.} {\bf 10} 11263 
\item Isern J Garc\'{\i}a--Berro E Hernanz M \& Chabrier G 2000 
      {\sl ApJ} {\bf 528} 397 
\item Isern J \& Garc\'{\i}a--Berro E 2008 {\sl Mem. SAI} {\bf 79} 545 
\item Isern J Garc\'{\i}a--Berro E Torres S \& Catal\'an S 2008 {\sl ApJ} 
      {\bf 682} L109
\item Isern J Hernanz M \& Garc\'{\i}a--Berro E 1992 {\sl ApJ} {\bf 392} L23
\item Kim J 1979 {\sl Phys. Rev. Lett.} {\bf 43} 103 
\item Knox R A Hawkins M R S \& Hambly N C 1999 {\sl MNRAS} {\bf 306} 736
\item Legget S K Ruiz M T \& Bergeron P 1998 {\sl ApJ} {\bf 497} 294
\item Liebert J Dahn C C \& Monet D G 1988 {\sl ApJ} {\bf 332} 891
\item Nakagawa M Kohyama Y \& Itoh N 1987 {\sl ApJ} {\bf 322} 291 
\item Nakagawa M Adachi T Kohyama Y \& Itoh N 1988 {\sl ApJ} {\bf 326} 241
\item Oswalt T D Smith J A Wood M A \& Hintzen P 1996 {\sl Nature} {\bf 382} 
      692
\item Peccei R D \& Quinn H R 1977a {\sl Phys. Rev. D} {\bf 16} 1791
\item Peccei R D \& Quinn H R 1977b {\sl Phys. Rev. Lett.} {\bf 38} 1440
\item Raffelt G G 2007 {\sl J. Phys. A: Math. Theor.} {\bf 40} 6607
\item Shifman M A Vainshtein A I \& Zakharov V I 1980 {\it Nucl. Phys. B} 
      {\bf 166} 43 
\item Salaris M Dom\'{\i}nguez I Garc\'{\i}a--Berro E Hernanz M Isern J \& 
      Mochkovitch R  1997 {\sl ApJ} {\bf 486} 413
\item Salaris M Garc\'{\i}a--Berro E Hernanz M Isern J \& Saumon D 2000 
      {\sl ApJ} {\bf 544} 1036 
\item Weidemann V 1968 {\sl ARAA} {\bf 6} 351
\item Weimberg S 1978 {\sl Phys. Rev. Lett.} {\bf 40} 223
\item Wilczek F 1978 {\it Phys. Rev. Lett.} {\bf 40} 279
\item Zhitnitskii A P 1980 {\it Sov. J. Nucl.} {\bf 31} 260 
\end{thereferences}

\end{document}